\documentclass[twocolumn,aps,prl,groupedaddress,showpacs, superscriptaddress]{revtex4-2} 

\usepackage{amsmath}
\usepackage{amssymb}
\usepackage{txfonts}
\usepackage{pxfonts}
\usepackage{gensymb}
\usepackage{graphicx,bm,units,yfonts}
\usepackage[table]{xcolor}
\usepackage{hyperref}
\usepackage{movie15}
\usepackage{xcolor}

\DeclareUnicodeCharacter{2212}{-}

\begin{document}

\title{Symmetry-Preserving Coupling Method for Topological Acoustic Metamaterials}

\author{Ssu-Ying Chen}
\email{sc945@njit.edu}
\affiliation{Department of Physics, New Jersey Institute of Technology, Newark, NJ, USA}

\author{Camelia Prodan}
\email{cprodan@njit.edu}
\affiliation{Department of Physics, New Jersey Institute of Technology, Newark, NJ, USA}

\date{\today}

\begin{abstract}

In this paper we investigate different types of couplings used in acoustic metamaterials requiring preservation of symmetries. For testing we use the SSH model to test whether topologically edge and interface modes are supported with the different types of connection. We observed that a modular platform where the resonators are coupled through the bottom is the simplest method that is accurate and flexible. 
\end{abstract}

\maketitle

\section{I. INTRODUCTION}



Topology opened a gate to realize new mechanical and acoustic systems. In recent years, phononic crystals have grasped the attention for offering inspiring opportunities to manipulate sound in new and unanticipated ways based on topological concepts.

Periodic acoustic systems, such as the topological boundary states based on the analogue of quantum Hall effect\cite{khanikaev2015topologically, yang2015topological}, the analogue of quantum spin Hall effect\cite{he2016acoustic}, the Floquet topological insulator\cite{peng2016experimental}, and the valley Hall effect\cite{lu2016valley} have been successively proposed and experimentally verified. Besides periodic systems, topological systems that lack periodicity can also achieve topological edge states, such as topological quasi-crystals \cite{ni2019observation, cheng2020experimental, apigo2019observation} where the topological structure is caused by disorder. The idea of topological acoustic insulators provides new schemes for designing devices with advanced functionalities. For example, the potential improvement of leaky-wave acoustic antenna\cite{Fleury2016}, directional topological acoustic antenna controlling sound for versatile applications\cite{Zhang2018}, specific signal filtering achieved by adding random disorder to clean structures. \cite{Zangeneh-Nejad2020}, various sound proof strategies\cite{Yamamoto2009, Xu2020, Liu2021}, and for a growing discussion of different fractal geometries.\cite{Song2016, Zhao2018, Singh2022} In essence, researchers have been and continue to intensely explored various topological acoustic systems, and among these researches, topological acoustic meta-material consisting of numerous phononic crystals is one of the focuses.

\begin{figure*}[ht!] 
\center
\includegraphics[width=\linewidth]{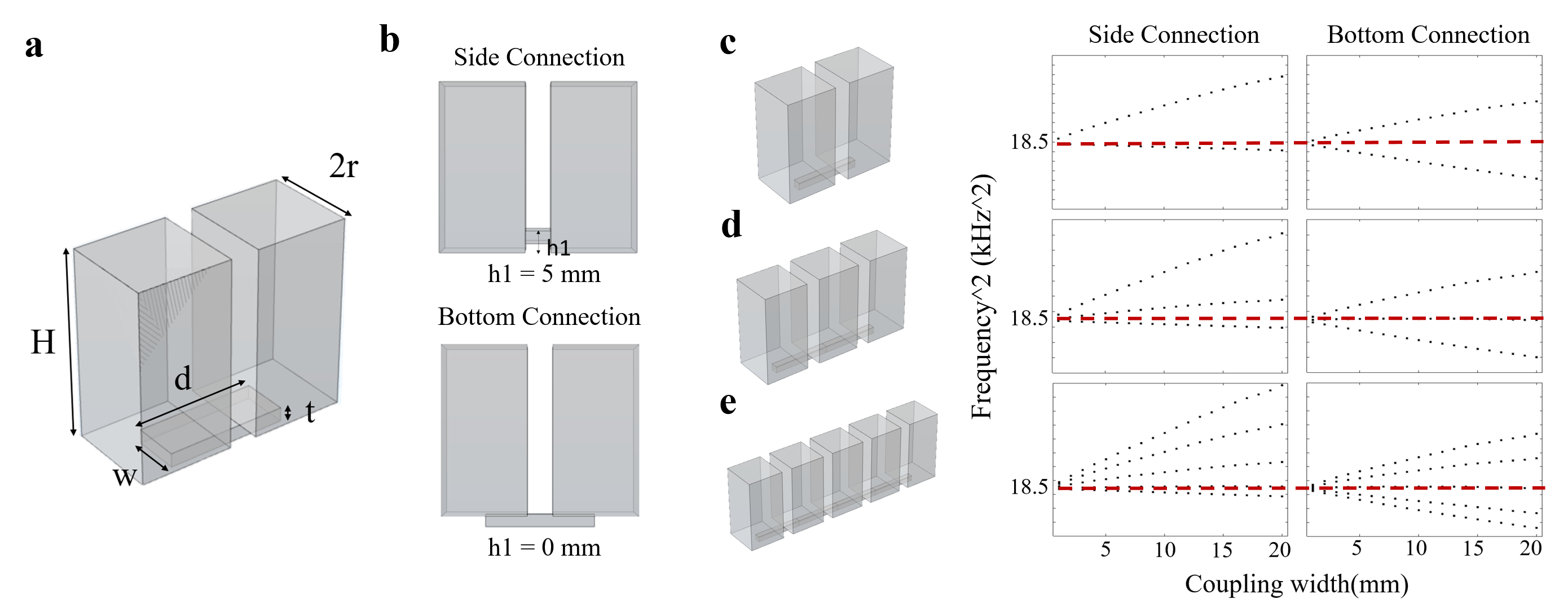}
\caption{\small {\bf Simulations showing that the bottom connection preserves the symmetry.} {\bf a} Dimensions of the resonators and coupling bridges. H = 40 mm, r = 5 mm. The coupling bridge has the width = 5 mm, the length d = 26 mm, the thickness t = 3 mm. {\bf b} The height position of the coupling bridge h1 = 5 mm for side connection, h1 = 0 mm for bottom connection. {\bf c} Dimer structure and the band spectrum of both side and bottom coupling methods. {\bf d} Trimer structure and the band spectrum of both side and bottom coupling methods. {\bf e} Pentamer structure and the band spectrum of both side and bottom coupling methods. It is clear in the band spectrum that in all three cases, the symmetry is preserved when the resonators are connected through bottom. Red dash lines indicates the symmetrical axis.}
\label{figure1}
\end{figure*} 


In view of the fact that topological insulators are also called symmetry-protected topological phases of matter, these gapped phases have topological properties relying on the presence of symmetries. This result is from a global property of the topological insulator's band structure: local perturbations cannot alter or damage this surface state. Since the properties of topological insulators and their surface states highly depend on the dimension of the material and its symmetries, they can be categorized using the periodic table of topological insulators\cite{Chiu2016}. From the experimental perspective, it is crucial to satisfy certain structures and to display the symmetry, moreover, preserving the symmetry is equivalent to preserving the topological properties. To preserve symmetry, coupling methods for topological acoustic experiments must be carefully chosen. The following question then arises: in discrete acoustic resonant models, how can the resonators be efficiently coupled to form a desired structure while preserving the symmetry so that it complies with the idea of topological metamaterial?



Some methods  preserve the symmetry using double side connection, meaning they added a coupling bridge that is also on the side and made the structure symmetric\cite{Zheng2022, Xiao2017, Ma2019, Qi2020}. The resonators are 3D printed  using photosensitive resins or other types of materials as coupling bridges must be fabricated with the resonators which makes as manufacture more complicated and time consuming. On top of that, if the dimension of the coupling bridge has to be changed, everything has to be made from scratch all over again.
 

In this article, we test an acoustic coupling method to connect the resonators through the bottom to preserve the symmetry and create a simple, flexible and Lego-like platform. We will compare two types of coupling methods throughout the paper: side coupling and bottom coupling(Fig.~\ref{figure1}{\bf b}). We will start by showing that bottom coupling preserves symmetry in simple periodic acoustic models while single side coupling fails to do so. Via the simulations of resonant modes of dimer, trimer, pentamer, and a classic Su–Schrieffer–Heeger model (SSH model) consisting of 14 resonators, as well as another SSH model with 28 resonators and a domain boundary. Furthermore, we address the corresponding experimental results of the dimer, the SSH model, and the SSH model with a domain boundary. All experimental results clearly show great agreement to the simulations, proving the effectiveness and simplicity of bottom coupling method.
\vspace{1mm}
\section{II. RESULTS and DISCUSSION}

The dimensions of the resonators that were used throughout the simulations and experiments are shown in (Fig.~\ref{figure1}{\bf a,b}).

\subsection{Side Connection vs. Bottom Connection}

To demonstrate and to compare the effect of side connection and bottom connection for topological acoustic resonators, numerical simulations were done using COMSOL Multiphysics software.

The simulated results is reported in Fig.~\ref{figure1} where the structures of dimer, trimer and pentamer of resonators with both side bridge coupling and bottom bridge coupling (Fig.~\ref{figure1}{\bf b}). The band spectrum (Fig.~\ref{figure1}{\bf c,d,e}) were generated by sweeping the width of coupling bridge from 1~mm to 20~mm with 1~mm steps and plotting corresponding eigenfrequencies versus widths. The wider coupling bridge means stronger coupling strength. In the band spectrum, the second mode split and the gap became larger when the coupling strength went up. With red dash lines as symmetry axes, it is clear that with side connection, the splitting modes are not symmetrical. Once we moved the coupling bridges to the bottom, the symmetry was restored.

\subsection{Dimer experiment}

\begin{figure*}[ht!]
\center
\includegraphics[width=\linewidth]{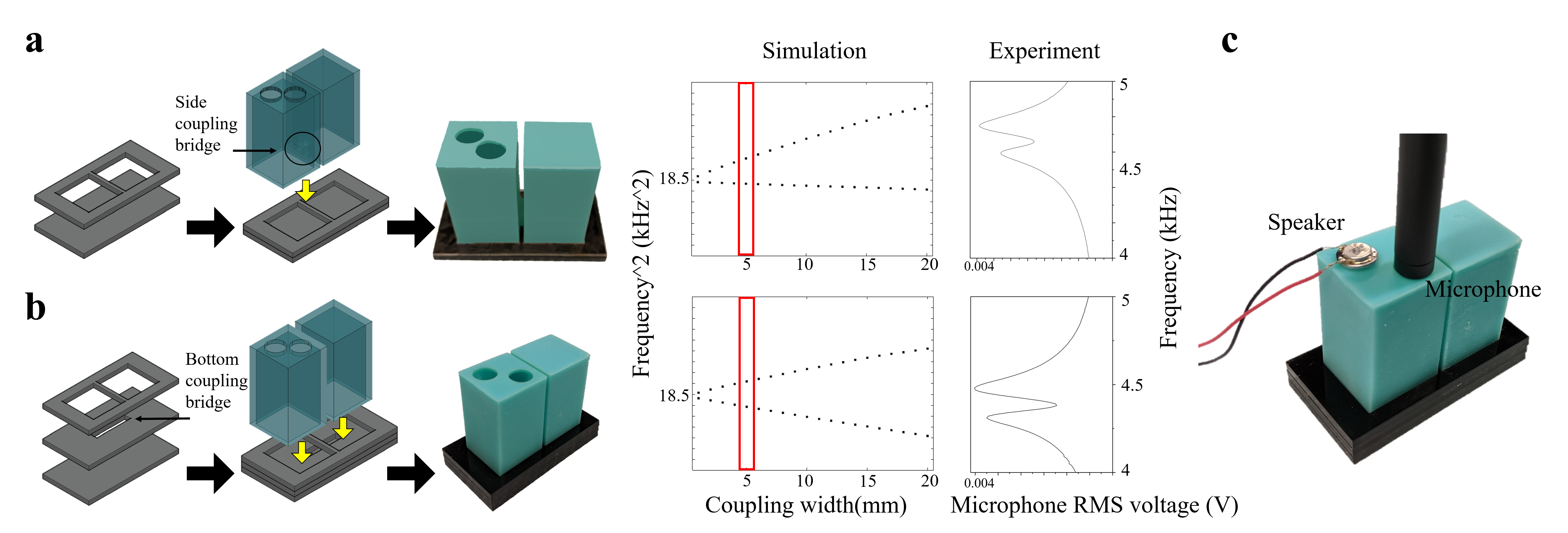}
\caption{\small {\bf Dimer experimental setups as well as the comparison of simulation and experiment results.} {\bf a} Assembly process of dimer connected through side. The dimer was 3D-printed together with a side coupling bridge, the height position of the top of the side coupling bridge h1 is equal to 7~mm. {\bf b} Assembly process of dimer connected through bottom. Two resonators were 3D-printed separately, and the bottom coupling bridge is grooved in the middle acrylic sheet as depicted. Red boxes in both band spectrum indicate the coupling width (5~mm) used in the experiment. {\bf c} The experimental setup.}
\label{figure2}
\end{figure*}

The experiment of dimer coupling was done to confirm the simulation results. The assembly process is shown in Fig.~\ref{figure2}{\bf a}, and the experimental set up is shown in Fig.~\ref{figure2}{\bf c}. The speaker and the microphone were placed on top of the same resonator to give off and collect sound, respectively. The frequency of the input signal sent to the speaker was swept from 4~kHz to 5~kHz in intervals of 10~Hz. Experimental results of dimer with bottom coupling bridges show a great agreement to the simulations (Fig.~\ref{figure2}{\bf b}). The middle of 2 peaks is about 4.3~kHz, same as where the 2 modes start to split in the band spectrum (around 18.5~$kHz^2$). However, with the side connection, the peaks shifted from the spectrum. Additionally, the height difference in peaks for bottom coupling is noticeably smaller than that of the side coupling (Fig.~\ref{figure2}{\bf b}). Comparing with the results from bottom coupling method, the symmetry was distinctly absent for side coupling.

\subsection{SSH acoustic model}

\begin{figure*}[ht!]
\center
\includegraphics[width=\linewidth]{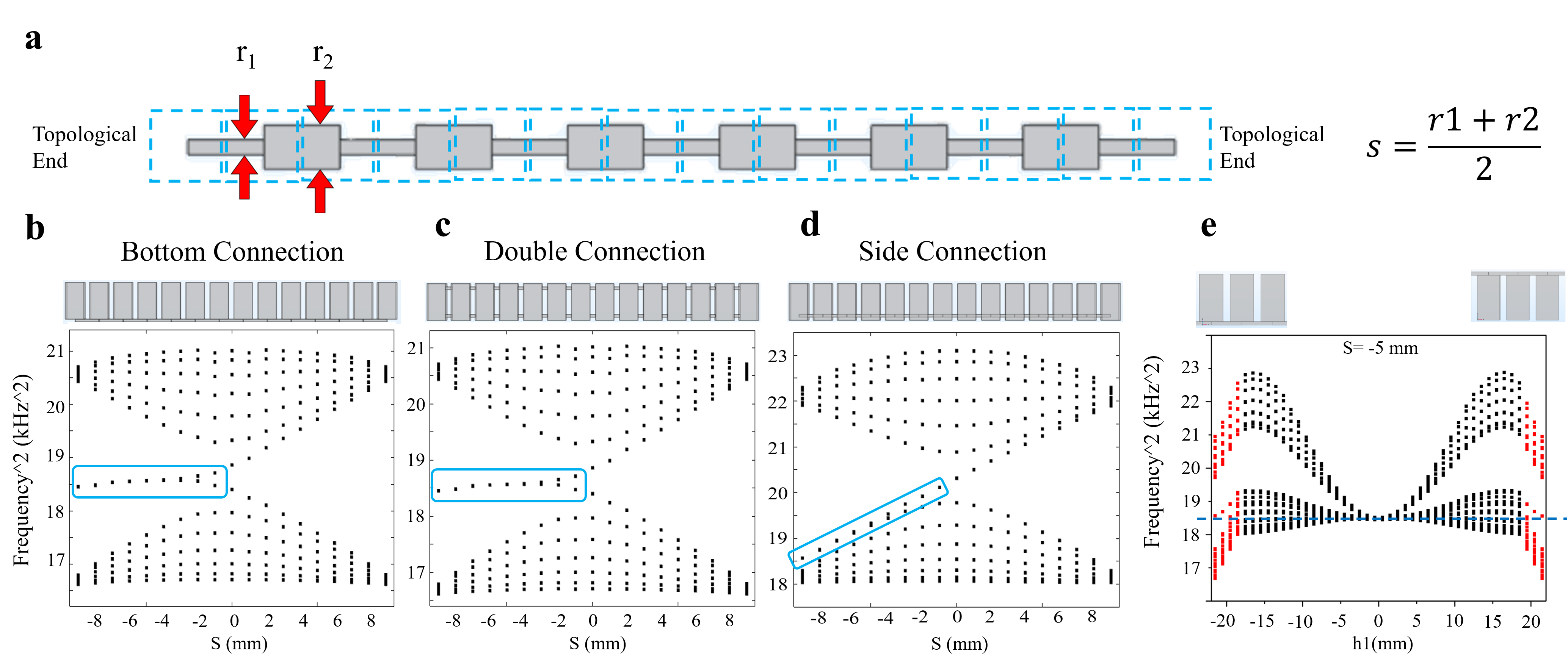}
\caption{\small {\bf Band spectrum of different types of connection in SSH model of 14 resonators, sweeping S. S is half of the difference in widths between strong and weak coupling bridge.} {\bf a} Top view of SSH coupling bridges. $r_{\textrm{1}}$ and $r_{\textrm{2}}$ are related to the alternate coupling strength. Light blue dash lines are where the resonators were placed. {\bf b} SSH model with bottom connection. {\bf c} SSH model with double connection from the sides. {\bf b} and {\bf c} has very similar spectrum. Blue boxes includes edge resonant modes. {\bf d} SSH model with side connection, the edge band merges into bulk. {\bf e} Band spectrum of SSH model sweeping the height of the coupling bridges. Red parts represent when the coupling bridges protrude from the bottom or top. Dark blue dash line marks 18.5~$kHz^2$.}
\label{figure3}
\end{figure*}

To further explore how the position of connection between acoustic crystals would influence topological gaps in band spectrum, we started by simulating 3 types of connection for SSH model of 14 resonators. The first type is to couple them with bridges connected through side, the second one is double side coupling bridges, and the third one is bottom connection. Fig.~\ref{figure3}{\bf a} shows the top view of SSH coupling bridges. $r_{\textrm{1}}$ and $r_{\textrm{2}}$ are the widths of coupling bridges, therefore they are related to the alternate coupling strength. Light blue dash lines indicate where the resonators were placed. The geometries of 3 coupling types are shown in Fig.~\ref{figure3}{\bf b, c, d}. The resonant spectrum were generated by sweeping S -- half of the difference in widths between strong and weak coupling bridge -- from -9~mm to 9~mm. When S equals to zero, all coupling bridges have an identical width of 10~mm.

As one can see in the spectrum (Fig.~\ref{figure3}{\bf b, c, d}, there is no gap for a uniform coupling connection (S = 0~mm, $r_{\textrm{1}}$ = $r_{\textrm{2}}$ = 10~mm). A bulk spectral gap opens once the connecting channels are set in an alternating strong/weak coupling strength ($S \ne 0~mm$). Furthermore, the bulk spectrum remains symmetric with respect to the middle of the bulk spectral gap (the small deviations are less than 5\% when compared with the overall width of the bulk spectrum). 

In (Fig.~\ref{figure3}{\bf b}), SSH model with bottom connection, one can observe the expected edge resonant modes, whose energies are pinned in the middle of the bulk spectral gap around 18.5~$kHz^2$, where the second mode of resonant frequency of a single resonator is. We can say that SSH model with bottom connection, COMSOL-simulated spectra displays an exact chiral symmetry. In Fig.~\ref{figure3}{\bf c}, SSH model with double side coupling bridges gives the same spectra, meaning both types of coupling approaches achieve the goal of protecting the chiral symmetry.

On the other hand, the spectra of SSH model connected through sides does not display the same symmetry. In Fig.~\ref{figure3}{\bf d}, not only do the edge modes merge into lower bulk bands, but the middle point of the spectrum shifted to around 2~$kHz^2$, demonstrating the same impracticality as in dimer experiments.

We also checked to see if the height position (h1 in Fig.~\ref{figure1}{\bf b}) of side coupling plays a role in preserving the symmetry. The height of side coupling bridge was swept from bottom to top to generate the spectra in Fig.~\ref{figure3}{\bf e}, while the coupling width were fixed at $r_{\textrm{1}}=15$ and $r_{\textrm{2}}=5$ (S = -5~mm). When h1 = -21.5 and 21.5~mm, the resonators are coupled from the bottom or from the top which resulted in bulk modes symmetric with respect to 18.5~$kHz^2$, and the edge modes appear at middle. Red parts represent when the coupling bridges protrude from the bottom or the top, meaning part of them are outside of the resonators, there are still edge modes existing in the gap but the symmetry shifts further and further away from the dash line. One can see that once the coupling bridges enter the body of resonators, the edge modes merge into bulk and the symmetry disappears.

\begin{figure*}[ht!]
\center
\includegraphics[width=\linewidth]{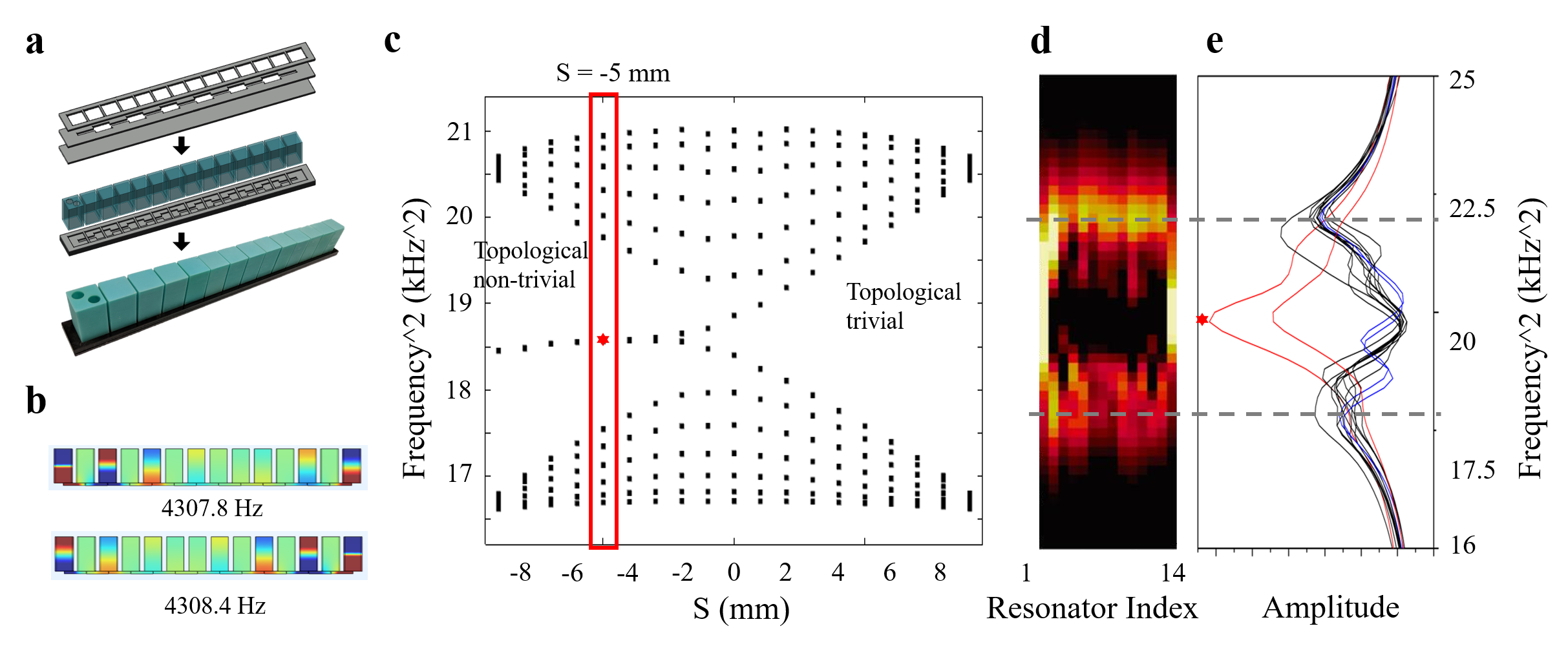}
\caption{\small {\bf SSH experiments demonstrate the simplicity and efficiency of bottom coupling method.} {\bf a} The assembling process of SSH model of 14 resonators connected through bottom. {\bf b} Acoustic pressure field distribution for the edge modes marked as red dot in panel {\bf c} when S = -5~mm. {\bf c} COMSOL simulated SSH model resonant spectrum. The red vertical box indicates S = -5~mm which makes $r_{\textrm{1}}=15$ mm and $r_{\textrm{2}}=5$ mm, the parameters were used in the experiments. {\bf d} Experimentally measured local density of states, assembled from normalized microphone readings from the top of the block resonators. The bright dispersive modes indicates the bulk and edge modes. {\bf e} Collapse on the frequency axis of the intensity plot in {\bf d}. The spectral gap is clearly recognized and the edge modes that show up in the gap are marked with a red star.}
\label{figure4}
\end{figure*}

Next, we did experiments with SSH model connected through bottom to confirm if the symmetry is also seen in the experimental results. Fig.~\ref{figure4}{\bf a} shows how the setup was made. The COMSOL-simulated acoustic pressure fields of the edge resonant modes are shown in Fig.~\ref{figure4}{\bf b}. The simulated band structure in Fig.~\ref{figure4}{\bf c} is added as a reference (same as Fig.~\ref{figure3}{\bf b}).

In these measurements, same as dimer experiments, the speaker and the microphone were inserted in the same resonator via two holes open at the top and the speaker's frequency was swept from 4 kHz to 5 kHz and the microphone picked up corresponding signals. The measurements were repeated for all resonators and the collected data was assembled in the local density of states plot shown in Fig.~\ref{figure4}{\bf d}. It is worth mentioning here that all resonators are removable and interchangeable, so the one with holes can be placed at any probe position desired for measurements. Panel {\bf e} provides an alternative depiction of the same data. The spectral gap as well as the expected edge and interface modes can be clearly identified and they are well aligned with the simulation in panel {\bf c}.

The density of states reported in Fig.~\ref{figure4}{\bf d} was obtained by integrating the local density of states acquired from resonators whose index are the same as the position of the probe. The same instrumentation was used. The measurements were repeated while moving the position of the probe. For each measurement, the frequency was scanned from 4~kHz to 5~kHz in 20~Hz steps.

Finally, we test and compare two coupling methods in a SSH model with a domain boundary(interface), both simulationally and experimentally. A SSH model with a domain boundary separates two topologically distinct SSH insulating phases, with one non-trivial edge(left) and one trivial(right). It is shown in Fig.~\ref{figure5}{\bf a} where $r_{\textrm{1}}=15$ mm and $r_{\textrm{2}}=5$ mm. Fig.~\ref{figure5}{\bf b} and Fig.~\ref{figure5}{\bf c} show the band spectrum of side connection and bottom connection, respectively. Red vertical boxes include resonant modes when S = -5~mm. Red and blue stars label where interface mode and edge modes appear, and the acoustic pressure field maps are shown below. Similar to the previous results, with bottom coupling, both simulation and experiment verified that the system contains topological resonant modes at the non-trivial edge as well as the domain boundary as expected, and they are located in the middle of the bulk band gap. In opposition, with side coupling, the edge mode disappears and the interface mode is close to the bulk, besides, it is noticeable that the energy is not as concentrated at interface from the acoustic pressure field distribution.

We also did experiments for bottom-connected SSH model with a domain boundary, the same protocol was applied and frequency was swept from 4 kHz to 5 kHz. The measurements were repeated for all 28 resonators and the collected data was then used to plot the local density of states which is shown in Fig.~\ref{figure5}{\bf d}. Panel {\bf e} shows the collapse of the data in panel {\bf d} on the frequency axis. The spectral gap as well as the expected edge modes can be clearly identified and are well aligned with the simulation in panel {\bf c}. The COMSOL-simulated acoustic pressure field maps of the edge resonant modes are shown in Fig.~\ref{figure5}{\bf b, c}.

\section{III. CONCLUSION}

In this paper we investigate acoustic coupling that preserve symmetries through Comsol simulations and experiments. We tested couplings through the bottom, and with bridges at different height through the side. The first test was done on dimmers, then on single SSH set up, as well as one with interface. We observed that the simplest way for such coupling is to connect the resonators through the bottom. Coupling through the side requires a second connection. 
The advantages of coupling through the bottom is the modular structure of the set up. The resonators and coupling are printed individually, allowing for an easy variation in the coupling strength when the experiment requires. This platform has an exceptional flexibility since the resonators can be stored for further use.

\section{IV. MATERIAL and METHODS}

\subsection{Simulation}

The simulations reported in all figures were performed with the COMSOL Multiphysics pressure acoustic module. The wave propagation domain shown in Fig.~\ref{figure1} was filled with air with a mass density 1.3~kg/m$^{3}$ and the sound's speed was set at 343 m/s, which is appropriate for room temperature. We shall consider the 3D printing UV resin material as hard boundary because of the huge acoustic impedance mismatch comparing with air.

\subsection{Experiment}

The resonators were 3D-printed using Anycubic Photon 3D printer, which uses UV resin and has 47~um XY-resolution and 10~um Z-resolution. The thickness of the walls is 2~mm, which ensures a high Q factor and justifies the rigid boundaries in the simulations. The inner dimensions of the resonators are shown in Fig.~\ref{figure1}a.

A dimer with a coupling bridge on the sides is 3D-printed as a whole. The width, length and the position of the coupling bridge are as labeled in Fig.~\ref{figure1}{\bf a}. One side was left open for ethanol rinsing and UV-curing. The dimer was then placed on a base of two layers of acrylic plates (top layer: 2~mm in thickness, bottom layer: 3~mm in thickness) to create a closed space for wave propagating(Fig.~\ref{figure2}{\bf a}). The reason that the top layer was 2~mm is to accommodate the side bridge. The bottom connection is achieved by assembling the supporting base, which consists of three layers of 3~mm thick acrylic plates(Fig.~\ref{figure2}{\bf b}). The middle layer of a groove is to account for acoustic coupling. The acrylic plates with patterns of the supporting bases were cut by the Boss Laser-1630 Laser Engraver. The nominal tolerance of the laser-cutter is 250~um.

\begin{figure*}[ht!]
\center
\includegraphics[width=\linewidth]{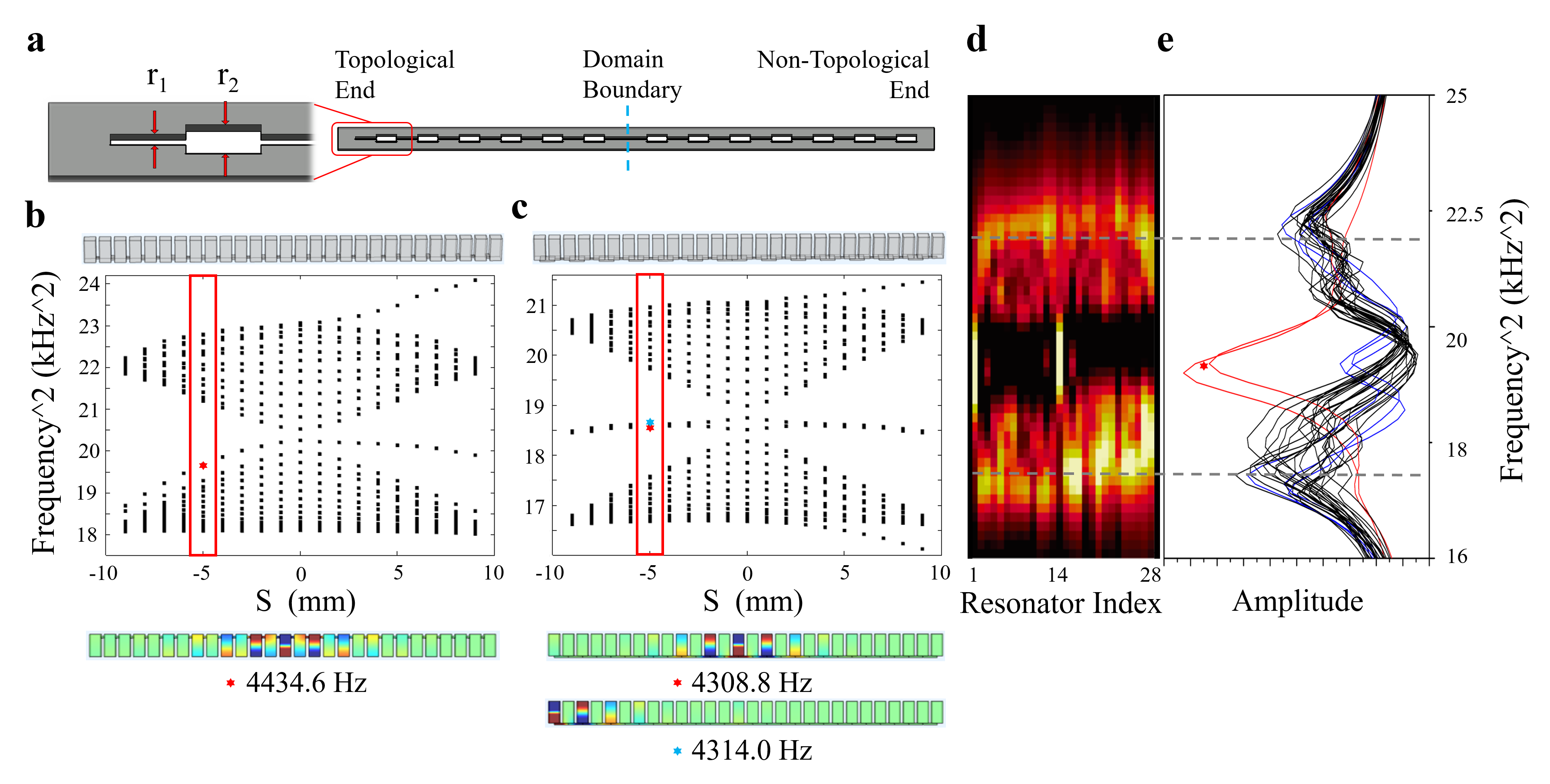}
\caption{\small {\bf The SSH model consists of 28 resonators with a domain boundary in the middle, topologically non-trivial at one end and trivial at the other.} {\bf a} Coupling bridges and its dimension. $r_{\textrm{1}}=5$ mm and $r_{\textrm{2}}=15$ mm. {\bf b} SSH model with side connection. The acoustic pressure field map below shows the interface mode at the red star in the spectrum when $S=-5$ mm. The model lacks edge modes. {\bf c} SSH model with bottom connection. Band structure and acoustic pressure field maps of both edge mode and interface mode. Blue and Red stars in the panel mark edge mode and interface mode, respectively. {\bf d} Experimentally measured local density of states, assembled from normalized microphone readings from the top of the block resonators. The bright dispersive modes indicates the bulk and edge modes. {\bf e} Collapse on the frequency axis of the intensity plot in {\bf d}. The spectral gap can be clearly identified and the edge and interface modes show up in the gap marked with a red star.}
\label{figure5}
\end{figure*}

For the SSH model connected from bottom, the same method was utilized. 14 resonators were placed and coupled through the channels with alternating widths grooved in the acrylic plates of the base. These resonators are detachable and interchangeable so that they can be moved around, thus acoustic crystals with different probe positions can be generated, and the resonators can be taken apart, stored and reassembled for new projects or designs.

The protocol for the acoustic measurements shown in Fig.~\ref{figure2}, Fig.~\ref{figure4} and Fig.~\ref{figure5} was as follows: Sinusoidal signals of duration 1~s and amplitude of 0.5~V generated by a Rigol DG 1022 function generator were sent out to a speaker placed in a porthole opened on top of a resonator. A dbx RTA-M Reference Microphone with a Phantom Power was inserted in a porthole next to the previous one and was used to acquire the acoustic signals (Fig.~\ref{figure2}{\bf c}). The signals were then read by a custom LabVIEW code via National Instruments USB-6122 data acquisition box and the data was stored for graphic renderings.

\section{ACKNOWLEDGMENT}
The authors acknowledges support from the National Science Foundation, grant CMMI-2131759.

\end{document}